\documentstyle[12pt]{article}
\begin{document}
\newcommand{\p}[1]{(\ref{#1})}
\newcommand{\be}{\begin{equation}}
\newcommand{\ee}{\end{equation}}
\newcommand{\nn}{\nonumber}
\newcommand{\sect}[1]{\setcounter{equation}{0}\section{#1}}
\renewcommand{\theequation}{\arabic{equation}}
\newcommand{\vs}[1]{\rule[- #1 mm]{0mm}{#1 mm}}
\newcommand{\hs}[1]{\hspace{#1mm}}
\newcommand{\mb}[1]{\hs{5}\mbox{#1}\hs{5}}
\newcommand{\Db}{{\overline D}}
\newcommand{\DDB}{\left[ D,\Db \right]}
\newcommand{\bea}{\begin{eqnarray}}
\newcommand{\eea}{\end{eqnarray}}
\newcommand{\PL}[1]{Phys.\ Lett.\ {\bf #1}}

\begin{titlepage}
\begin{flushright}
{\tt solv-int/9802018}\\
February 1998
\end{flushright}
\vskip 1.0truecm
\begin{center}
{\large \bf ON GAUGE-EQUIVALENT FORMULATIONS OF N=4 SKdV HIERARCHY}
\vglue 2cm
{\bf E. Ivanov}
\vglue 1cm
{\it Bogoliubov Laboratory of Theoretical Physics,
JINR,\\
141 980 Dubna, Moscow Region, Russia \\

\vspace{0.4cm}
e-mail: eivanov@thsun1.jinr.dubna.su}
\end{center}

\vspace{2cm}

\begin{abstract}
\noindent 
We point out that the $N=4$ supersymmetric KdV hierarchy, 
when written through the prepotentials of the bosonic chiral 
and antichiral $N=2$ supercurrents, exhibits a freedom related 
to the possibility to choose different gauges for the prepotentials. 
In particular, this implies that the Lax operator for the $N=4$ SKdV 
system and the associated realization of $N=4$ supersymmetry obtained in 
{\tt solv-int/9802003} are reduced to the previously known ones. 
We give the prepotential form of the `small' $N=4$ superconformal 
algebra, the second hamiltonian structure algebra of the $N=4$ SKdV 
hierarchy, for two choices of gauge. 
\end{abstract}

\end{titlepage}

\noindent{\bf 1.} $N=4$ supersymmetric KdV (SKdV) system with the `small' 
$N=4$ superconformal algebra (SCA) as the second hamiltonian structure 
was firstly constructed in \cite{1} in terms of harmonic $N=4$ superfields. 
Later on, it was 
reformulated in terms of $N=2$ superfields \cite{2} that allowed 
to prove its integrability (the existence of $N=4$ SKdV 
hierarchy) by constructing the relevant superfield Lax 
representations. Two different $N=2$ superfield 
Lax operators for $N=4$ SKdV were proposed \cite{3,4}. The crucial 
property of one of them is the manifest preservation of 
$N=2$ chirality \cite{3}, while another one \cite{4} is 
characterized by the commutativity with one of $N=2$ covariant spinor 
derivatives \cite{5} 
\footnote{The relation between these two operators is discussed 
in \cite{KS}.}. 

In a recent preprint \cite{7} it was claimed that there exists one more 
Lax operator for $N=4$ SKdV hierarchy, possessing the same commutativity 
property as that of \cite{4}. One of the aims of the present note is to 
demonstrate that these Lax operators are in fact identical to one another. 
The seeming difference between them is due to different, 
though gauge-equivalent, representations of two of the basic superfields 
of $N=4$ SKdV system, namely the spin 1 chiral and antichiral superfields, 
through their spin $1/2$ spinor prepotentials. In ref. \cite{4} 
the prepotentials were chosen to be chiral and antichiral $N=2$ 
superfields, while in \cite{7} there was used another gauge in which the 
prepotentials coincide and are given by a general $N=2$ superfield. 
We explicitly give the gauge transformation relating these two gauges and 
show that the realization of $N=4$ supersymmetry declared in \cite{7} 
to be new is gauge-equivalent to that given in \cite{4}. We also 
present the realization of classical `small' $N=4$ SCA on the spin $1/2$ 
prepotentials in both gauges. It proves to be non-local as distinct 
from the well-known realization in terms of the spin 1 supercurrents.    

\vspace{0.3cm}
\noindent{\bf 2.} We start by recalling the salient features of 
the Lax representation for $N=2$ SKdV hierarchy given in \cite{4}. 

We use the standard notation for the 
covariant derivatives of $N=2$ superspace 
$Z\equiv \{ z, \theta, \overline \theta \}$  
\bea
&& D = \frac{\partial}{\partial\theta}-
 \frac{1}{2}\overline \theta\frac{\partial}{\partial z},\quad 
\Db=\frac{\partial}{\partial\overline \theta}-
 \frac{1}{2}\theta\frac{\partial}{\partial z}~, \nn \\
&& \left\{ D,\Db \right\} = -\frac{\partial}{\partial z}~, \qquad
D^2 = \Db^2 =0\;. \label{not} 
\eea  

The $N=4$ SKdV Lax operator of ref.\cite{4} and the $N=4$ SKdV hierarchy 
equations, in the form adapted for 
our purpose, read 
\bea
&& L = \partial -J- \Db\partial^{-1}(DJ) -
F \Db\partial^{-1}(D\overline{F}) -
\Db\partial^{-1}(F D{\overline F})~, \label{lax1} \\
&& \frac{\partial}{\partial t_k} L= - \left[ (L^k)_{\geq1},L\right]\;,
\label{laxeq}
\eea
where the subscript $\geq 1$ marks the strictly differential 
part of pseudo-differential operator. In eq. \p{lax1}, 
the parentheses mean that the differential operator inside them acts only 
on the expression inside; otherwise, the operator is assumed 
to act freely to the right \footnote{This convention 
applies only to the Lax operators and to the operators in the r.h.s. of 
the Poisson brackets defining the second hamiltonian 
structure algebra; in all other cases (e.g., in the 
evolution equations) any operator (including $\partial^{-1}$) 
is assumed to act only on the function standing to its right.}. 
The general spin 1 $N=2$ superfield J(Z)
and the chiral spin $1/2$ superfields $F(Z)$, $\overline{F}(Z)$, 
\be
DF = \Db\; \overline F = 0~, \label{chir1}
\ee
are the basic objects of $N=4$ SKdV hierarchy in this $N=2$
superfield formulation. The spin 1 chiral supercurrents $\Phi(Z)$, 
$\overline{\Phi}(Z)$ which were used in \cite{1,2} and 
which, together with $J(Z)$, constitute the standard basis of 
the `small' $N=4$ SCA are related to $F, \overline F$ by 
\bea 
&&\Phi = \Db F~, \qquad \overline \Phi = D\overline F \label{def1phi} \\ 
&& \Db \Phi = 0~, \qquad D \overline \Phi = 0~. \label{chir2}
\eea
It is straightforward to check that $[D, L] = 0$.

As an example, we explicitly give the second flow equations
\bea
\frac{\partial J}{\partial t_2} & = & \left[ D,\Db \right] J{}'+ 2JJ{}'
- 2(\Db F D\overline F){}' =  \left[ D,\Db \right] J{}'+ 2JJ{}' - 
2(\Phi \overline \Phi){}'~, 
\label{Jflow2} \\
\frac{\partial F}{\partial t_2} & = &- {F}{}'' - 2D(J\Db F)~, 
\qquad 
\frac{\partial {\overline F}}{\partial t_2} = {{\overline F}}{}''-
2\Db (JD{\overline F})\;. 
\label{2flow1}
\eea
Acting on the last two equations by $\Db, D$ and making use of 
the relation \p{def1phi}, we can equivalently rewrite this system 
in terms of $\Phi$, $\overline \Phi$. 
The $\Phi, \overline \Phi$ form of eqs. \p{2flow1} is as follows  
\be
\frac{\partial \Phi}{\partial t_2} =  -\Phi {}''-2 \Db D \;(J\Phi)~, 
\qquad \frac{\partial \overline \Phi}{\partial t_2} =  \overline \Phi{}''-
2D \Db \; (J \overline \Phi)~. \label{phiflow2}
\ee
Similar equivalent forms can be given for other flows. This equivalence 
between the $F$ and $\Phi $ representations is based on the invertibility 
of the relation  \p{def1phi}: due to the chirality of $F, \overline F$ 
one can express the latter, up to an unessential constant Grassmann 
zero-mode, in terms of $\Phi, \overline \Phi$ 
\be \label{invers}
F = -\partial^{-1} D\Phi~, \qquad \overline F = -  \partial^{-1} 
\Db\; \overline \Phi~.
\ee

The $N=4$ SKdV hierarchy equations reveal covariance under an extra $N=2$ 
supersymmetry 
\be  \label{n4susy1}
\delta J = \epsilon D \Phi + \overline{\epsilon} \Db\; \overline \Phi\;, 
\;\;
\delta \Phi = -\overline{\epsilon} \Db J\;, \;\; 
\delta \overline \Phi = - \epsilon DJ\;.
\ee
Here, $\epsilon, \overline{\epsilon}$ are complex  
Grassmann parameters (they are not mutually conjugated in general). 
Together with the explicit $N=2$ 
supersymmetry these transformations constitute $N=4$ supersymmetry 
in one dimension. An equivalent realization in terms of 
the superfields $F, \overline F$ is non-local  
\be  \label{n4susyF}
\delta J = - \epsilon F{}'  
- \overline{\epsilon} \overline F{}'\;, \;\; 
\delta F = - \overline{\epsilon} D\partial^{-1} \Db J\;, \;\; 
\delta \overline F = - \epsilon \Db \partial^{-1} DJ\;.
\ee
It is easy to directly check the covariance of the second flow equations 
\p{Jflow2} - \p{phiflow2} under the transformations \p{n4susy1} or 
\p{n4susyF}.

Note that here and in what follows we do not need to impose any reality 
conditions on the involved superfields.

\vspace{0.3cm}
\noindent{\bf 3.} It has been noticed in \cite{4} that 
the superfields $F, \overline F$ can be considered as prepotentials of 
$\Phi, \overline \Phi$ in some special gauge. Let us elaborate  
on this interpretation in more detail and show that the description 
of $N=4$ SKdV hierarchy in terms of $J(Z)$ and a general single spinor 
superfield $g(Z)$ proposed in \cite{7} merely amounts to another 
choice of gauge for the prepotential. 

A general solution of the chirality conditions \p{chir2} is as follows 
\be  \label{gensol}
\Phi = \Db v~, \qquad \overline \Phi = D\overline v~, 
\ee
with $v, \overline v $ being general complex spin $1/2$ fermionic 
prepotentials (they, like $\Phi$ and $\overline \Phi $, are not 
obliged to be mutually conjugated). The prepotentials are defined up 
to gauge transformation 
\be  \label{gauge1}
v \Rightarrow v + \Db \lambda~, \qquad 
\bar v \Rightarrow \overline v - D \overline \lambda~,  
\ee
where $\lambda $, $\overline\lambda $ are arbitrary complex bosonic 
superfield parameters. 
In turn, the latter are defined up to the freedom 
\be  \label{gauge2}
\lambda \Rightarrow \lambda + \Db \omega~, \quad 
\overline \lambda \Rightarrow \overline \lambda + D \overline \omega
\ee
$\omega $, $\overline\omega $ being complex fermionic superfield 
functions.

One can make use of the gauge freedom \p{gauge1} to choose different gauges 
for the prepotentials $v$, $\overline v$. Let us first demonstrate 
that the choice 
\be
v_1 = F~, \qquad \overline{v}_1 = \overline F~, \qquad DF = 
\Db\;\overline F = 0 
\label{choice1}
\ee
is just one of such possible gauge-fixings. We start from a general $v$ 
and should show the existence of a gauge function $\lambda_1$ such that 
\be 
F = v + \Db \lambda_1~.
\ee
Using the algebra of covariant derivatives \p{not}, it is easy 
to check that this condition fixes $\Db \lambda_1$ up to an 
unessential complex Grassmann constant (it can be absorbed into $F$)
\be
\Db \lambda_1 = \partial^{-1}\Db D v \quad \Rightarrow \quad
\lambda_1 = \partial^{-1} D v  + \Db \omega~, 
\ee
whence 
\be
F = v + \partial^{-1} \Db D v~.
\ee
For $\overline F$ one obtains an analogous expression in terms of 
$\overline v$.

Another possible gauge choice is to identify $v$ and $\overline v$ 
\be \label{choice2}
v_2 = \overline{v}_2 = g~.
\ee 
Let us show the existence of gauge transformation relating 
the gauges \p{choice1} and \p{choice2}
\be
F = g + \Db \lambda_2~, \qquad \overline F = g - D \overline \lambda_2~.
\ee
Once again, the use of the algebra \p{not} allows one 
to determine $\Db \lambda_2$, $D \overline \lambda_2$
up to unessential Grassmann constants in terms of either $g$ or 
$F, \overline F$. 
As the result we get the following simple invertible relations 
between $F, \overline F$ and $g$
\bea
&& F = g + \partial^{-1} \Db D g~, \qquad  
\overline F = g + \partial^{-1} D \Db g  \label{rel1} \\
&& g = F - \partial^{-1} \Db D \overline F = F + \overline F~. 
\label{rel2}
\eea

It should be stressed once more that the chiral supercurrents $\Phi $, 
$\overline \Phi $ are gauge-invariant quantities, i.e. they do not 
depend on 
the choice of gauge for $v$, $\bar v$. In other words, from the point 
of view of $N=4$ SKdV hierarchy it does not matter which kind of the 
prepotential representation has been chosen for these objects
\be \label{corresp}
\Phi = \Db F = \Db g~, \qquad \overline \Phi = D\overline F = Dg~. 
\ee  
The $F, \overline F$ representation was used in \cite{4}. It is just the 
$g$ representation that was used in \cite{7}. Obviously, even more choices 
are possible.

A common feature of the above gauges is that they fix the prepotentials 
up to a constant, i.e. allow no non-trivial residual gauge freedom. 
Respectively, the field component sets of the prepotentials 
and the supercurrents $\Phi, \overline \Phi$ precisely match each other in 
these gauges.
Just due to this property one can establish invertible relations between 
the gauge-fixed prepotentials and $\Phi $, $\overline \Phi$. 
Using the explicit relation \p{rel2}, it is easy, e.g., to find 
the analog of the relation \p{invers} for the gauge \p{choice2}
\be  \label{gphi}
g = -\partial^{-1}\left( D\Phi + \Db\; \overline \Phi \right)~.
\ee

With all these explicit relations at hand, it is straightforward 
to be convinced that all the equations of $N=4$ SKdV hierarchy 
in the $g$-description \cite{7} directly stem from those in the 
$F, \overline F$ description \cite{4}. E.g., the second flow 
equations \p{2flow1} 
entail, through the relation \p{rel2}, the following evolution 
equation for $g$ 
\be \label{2flowg}
{\partial g\over \partial t_2} = \left[ D, \Db \right] g{}' - 
2D\;(J\Db g) - 2\Db \;(JDg)
\ee
(it again implies \p{phiflow2} through the correspondence \p{corresp}).
The same is true for the $N=4$ supersymmetry transformations: 
eqs. \p{n4susyF} have the following gauge-equivalent form in terms of 
the superfields $J, g$ (one should use the relations \p{rel1} and \p{rel2})
\be
\delta J = \epsilon D\Db g + \overline \epsilon \Db Dg~, \quad 
\delta g = -\epsilon \partial^{-1}\Db D J - 
\overline \epsilon \partial^{-1} D\Db J~.
\ee

It is interesting to see how the chiral subvariety of the $N=4$ SKdV 
equations look in terms of $v, \overline v$ before any gauge fixing. 
For the second flow one gets from \p{phiflow2}, \p{gensol} (we explicitly 
write only the equation for $v$)
\be \label{2ndv}
\frac{\partial v}{\partial t_2} = - {v}{}'' - 2D(J \Phi) + \Db X~.
\ee 
Here the superfield $X(Z)$ describes an arbitrariness related 
to the gauge freedom \p{gauge1}. To different gauges for $v$, 
$\overline v$
there correspond special choices of $X$ (and $\overline X$). A 
similar arbitrariness persists in the higher-dimension flows 
written through $v, \overline v$.  

Let us now apply to the Lax operator \p{lax1}. Keeping in mind the 
invertible relations between $\Phi, \overline \Phi$ on the one hand and 
$\overline F, F$ or $g$ on the other, it is natural to expect that it 
admits a unique gauge-invariant representation in terms of the 
supercurrents $\Phi, \overline \Phi$. Then its $F, \overline F$ form 
\p{lax1} or a gauge-equivalent $g$ form should follow from this `master' 
representation upon substituting different expressions for 
$\Phi, \overline \Phi$ through the prepotentials $v, \overline v$.  This 
is indeed so. The $\Phi, \overline \Phi$ representation of \p{lax1} is 
as follows 
\be
L = \partial -J- \Db\partial^{-1}(DJ) - \partial^{-1}\Phi\overline \Phi 
+ \partial^{-1}(D\Phi) \partial^{-1} \Db\; \overline \Phi~. \label{lax2}
\ee
Substituting for $\Phi, \overline \Phi$ their expressions through  
$\overline F, F$ (eqs. \p{def1phi}), using the $D, \Db$ algebra \p{not} 
and the identities like 
$$
(F{}') = \partial F - F \partial~, 
$$
it is straightforward to show that \p{lax2} is identical to \p{lax1}. 
Of course, one can arrive at \p{lax2} by substituting 
\p{invers} in \p{lax1}. The $g$ form of the Lax operator can be obtained 
by substituting the relevant expressions for $\Phi, \overline \Phi$ from 
eqs. \p{corresp}. 

Comparing the operator \p{lax2} with the Lax operator proposed 
in \cite{7}, one observes them to be identical to each other up to  
the following redefinitions 
\be \label{red1}
\Phi = \overline G~, \qquad \overline \Phi = G~,   
\ee 
\be \label{replace}
G, \overline G \quad \Rightarrow \quad iG, i\overline G~ \qquad 
( F, \overline F 
\quad \Rightarrow \quad iF, i\overline F~, \qquad g \quad 
\Rightarrow \quad i g)~. 
\ee

Let us notice that the Lax operator \p{lax2} can be brought into the 
following suggestive form
\be \label{lsugg}
L = \partial - J -\Phi \partial^{-1} \overline \Phi - \Db \partial^{-1} 
\left[ D, \; J + \Phi \partial^{-1} \overline \Phi \right]~, 
\ee
from which it is easy, e.g., to reveal that on the subspace of the chiral 
wave functions $\Psi(Z)$, $D\Psi(Z) = 0$, this operator is reduced 
to that proposed in \cite{3}
\be 
L \Rightarrow -D\Db - D\Db\partial^{-1}\left\{J + \Phi\partial^{-1} 
\overline \Phi 
\right\} D\Db\partial^{-1}~.
\ee
The existence of such a correspondence for more general case was mentioned 
in \cite{KS}.

All the basic features of the prepotential formulation of $N=4$ SKdV 
hierarchy remain valid as well for another known $N=2$ hierarchy with 
the `small' $N=4$ SCA as the second hamiltonian structure, the so 
called `quasi' $N=4$ SKdV system \cite{DGI} (its characteristic feature 
is lacking of $N=4$ supersymmetry). In particular, it admits 
gauge-equivalent formulations in terms of the superfields 
$J, F, \overline F$ or $J, g$.

\vspace{0.3cm}     
\noindent{\bf 4.} As the last topic, we will give how the `small' 
$N=4$ SCA, the second hamiltonian structure algebra of $N=4$ SKdV 
hierarchy, is realized on the $N=2$ SCA supercurrent $J(Z)$ and the 
prepotentials $v(Z), \bar v(Z)$ in the gauges \p{choice1}, \p{choice2}. 

The classical non-vanishing Poisson brackets (PB) defining this algebra 
on the supercurrents $J, \Phi, \overline \Phi $ can be written as follows 
(see, e.g. \cite{IKT} \footnote{The precise correspondence with the  
PBs of ref. \cite{IKT} is achieved via substitutions $\Phi \leftrightarrow 
\overline \Phi~, \;\{ \} \rightarrow - \{ \}$.})
\bea
\{ J(1), J(2) \} &=& \left( \partial \left[ D, \Db \right] + \partial J 
+ J \partial + DJ \Db + \Db JD 
 \right)\delta(1,2)~,  \label{JJ} \\
\{J(1), \Phi(2)\} &=& -\left(\Db D \Phi + \Db \Phi D \right)\delta(1,2)~,
\label{Jphi} \\
\{J(1), \overline \Phi(2)\} &=& 
- \left(D \Db\; \overline \Phi + D \overline \Phi \Db \right) 
\delta(1,2)~, \label{Jbar} \\
\{\Phi(1), \overline \Phi(2) \} &=& \left( \partial D\Db + DJ\Db \right) 
\delta (1,2)~, \label{phibar}
\eea
where $\delta(1,2) = (\theta_1 - \theta_2)(\overline \theta_1 - 
\overline \theta_2) 
\delta (z_1 - z_2) $ and the differential operators in the r.h.s. are 
evaluated at the second point of $N=2$ superspace. Using the inverse
relations \p{invers}, it is straightforward to rewrite PBs 
\p{Jphi} - \p{phibar} in terms of $F, \overline F$
\bea 
\{J(1), F(2) \} &=& \left(F D \Db  - D\Db F + 
\partial^{-1} D\Db F \Db D \right)\delta (1,2)~, \label{JF} \\
\{J(1), \overline F(2) \} &=& \left(\overline F\; \Db D -\Db D 
\overline F +  
\partial^{-1} \Db D \overline F D \Db \right)\delta (1,2)~, 
\label{JbarF} \\
\{F(1), \overline F(2) \} &=& \left(\Db D - 
\partial^{-1}\Db D J \Db D\partial^{-1} \right)\delta(1,2)~. 
\label{FbarF} 
\eea 
Quite analogously, using the relations \p{gphi} or \p{rel2}, one easily 
restores from eqs. \p{Jphi} - \p{phibar} or 
\p{JF} - \p{FbarF} the $J, g$ form of the `small' $N=4$ SCA. It is given 
by the PB \p{JJ} and the following PBs involving $g(Z)$
\bea
\{J(1), g(2) \} &=& \left( \partial g + {1\over 2} g \partial 
- Dg\Db - \Db g D - {1\over 2}\partial^{-1}\left[ D,\Db \right] g 
\left[ D,\Db \right] \right) \delta(1,2)~, \label{Jg} \\
\{g(1), g(2) \} &=& - \left( \left[ D, \Db \right] + {1\over 2} J + 
{1\over 2}\partial^{-1}\left[ D,\Db \right] 
J \partial^{-1}\left[ D,\Db \right] \right) \delta(1,2)~. \label{gg}
\eea

An interesting peculiarity of the `small' $N=4$ SCA written in this way is 
the unavoidable presence of non-local terms in the r.h.s. of the 
defining PBs. In particular, the relations \p{JF}, \p{JbarF}, \p{Jg} 
imply that the fermionic superfields $F, \overline F$ and 
$g$ are not primary in the standard sense with respect to $N=2$ SCA 
(the notion of conformal spin is still meaningful). Note that the set 
of PBs \p{JJ}, \p{Jg}, \p{gg} differs from the PBs of $N=3$ SCA 
in the $N=2$ superfield notation \cite{7} just by such 
non-local terms. 

As a final remark, we note that it would be desirable to find an analog 
of the above PBs for the arbitrary prepotentials $v, \overline v$ 
before imposing any gauge condition on them. Such an algebra (if existing) 
is expected to be more general than the `small' $N=4$ SCA. The 
latter would follow from this general algebra as some its reduction 
upon imposing the appropriate gauge conditions and applying 
the Dirac procedure. Knowing the general algebra could 
also fix the above-mentioned uncertainty in the evolution 
equations for $v, \overline v$.

\vspace{0.5cm} 
\noindent{\bf Acknowledgement.}  We acknowledge a partial support from 
the grants RFBR-96-02-17634, RFBR-DFG-9602-00180, INTAS-93-127-ext and 
INTAS-96-0538.


\begin{thebibliography}{99}
\bibitem{1} F. Delduc and E. Ivanov,  Phys. Lett. {\bf B 309} (1993) 
312; {\tt hep-th/9301024}
\bibitem{2} F. Delduc, E. Ivanov, S. Krivonos, J. Math. Phys. 
{\bf 37} (1996) 1356, {\bf 38} (1997) 1224E; {\tt hep-th/9510033}
\bibitem{3} F. Delduc, L. Gallot, Commun. Math. Phys. {\bf 190} 
(1997) 395; {\tt solv-int/9609008}
\bibitem{4} E. Ivanov, S. Krivonos, Phys. Lett. {\bf A 231} (1997) 75; 
{\tt hep-th/9609191}
\bibitem{5} A. Sorin, {\it `The discrete symmetries of the $N=2$ 
supersymmetric GNLS hierarchies'}, JINR E2-97-37; {\tt solv-int/9701020} 
\bibitem{KS} S. Krivonos, A. Sorin, {\it `Extended $N=2$ supersymmetric 
matrix $(1,s)$-KdV hierarchies'}, JINR E2-97-365; {\tt solv-int/9712002}
\bibitem{7} S. Krivonos, A. Pashnev, Z. Popowicz, {\it `Lax pairs for 
$N=2,3$ supersymmetric KdV equations and their extensions'}, 
IFT UWr 919/98; {\tt solv-int/9802003}
\bibitem{DGI} F. Delduc, L. Gallot, E. Ivanov, Phys. Lett. {\bf B 396} 
(1997) 122; {\tt hep-th/9611033}
\bibitem{IKT} E. Ivanov, S. Krivonos, F. Toppan, Phys. Lett. 
{\bf B 405} (1997) 85; {\tt hep-th/9703224}
 
\end{thebibliography}
\end{document}